\documentclass[aps,amsmath,prx,reprint,superscriptaddress]{revtex4-1}

\usepackage{bm}
\usepackage{color}
\usepackage{soul}
\usepackage{graphicx}
\usepackage[normalem]{ulem}
\usepackage{amsmath}
\newcommand{\tna}{\mathbf{{a}}}
\newcommand{\tnb}{\mathbf{b}}
\newcommand{\abar}{\bar{\tna}}
\newcommand{\bbar}{\bar{\tnb}}
\newcommand{\Kbar}{\bar{K}}

\newcommand\degC{$^o$\,C}
\newcommand{\ie}{{\em i.e.,}~}
\newcommand{\eg}{{\em e.g.,}~}

\newcommand\be{\begin{equation}}
\newcommand\ee{\end{equation}}
\newcommand\bea{\begin{eqnarray}}
\newcommand\eea{\end{eqnarray}}

 \renewcommand{\hat}[1]{\oldhat{\mathbf{#1}}}

\let\oldhat\hat

\graphicspath{{Figures/}}
\begin{document}
\title{Oscillating membranes: modeling and controlling autonomous shape-transforming sheets}
\author{Ido Levin}
\affiliation{Racah Institute of Physics, The Hebrew University, Jerusalem 91904, Israel}
\author{Robert D. Deegan}
\affiliation{Department of Physics, University of Michigan,
Ann Arbor, Michigan 48109, USA}
\author{Eran Sharon}
\email[]{erans@mail.huji.ac.il}
\affiliation{Racah Institute of Physics, The Hebrew University, Jerusalem 91904, Israel}

\date{\today}

\begin{abstract}
Living organisms have mastered the dynamic control of internal stresses to perform an array of functions, such as change shape and locomote.
State-of-the-art attempts to replicate this ability in synthetic materials are rudimentary in comparison.
Here we present the first experimental realization of a self-oscillating gel in a thin sheet configuration.
We show that internal signaling produces stresses that drive lifelike shape changes, that the material’s response is accurately modelled with the theory of non-Euclidean elasticity and that the internal signaling can be programmed with light.
Together, our results demonstrate a complete route for developing fully autonomous soft machines.
\end{abstract}

\maketitle

\section{Introduction}
From an amoeba’s ability to extend pseudopods to the fast active tissue deformations in multi-cellular primitive animals to the muscle contractions of higher animals, nature provides countless examples of soft materials undergoing autonomous mechanical deformations~\cite{Nath2003,YONEDA1982,Armon2018}.
Nevertheless, state-of-the-art attempts to replicate these capabilities in synthetics systems are crude by comparison.
Typically, these rely on an external, discrete actuators, or global control schemes ~\cite{klein:2007a, Whitesides2018,Hajiesmaili2019}, and the results thus lack the autonomy, flexibility, and configurablity of living systems. 

To achieve the dynamics of living systems, an internal, spatially-varying actuation and control scheme is needed that capitalizes on the facility of soft materials to undergo large and complex deformations.
Self-oscillating gels~\cite{yoshida:1997a} are a promising class of materials for realizing these characteristics.
These gelatinous materials shrink and swell in response to the phase of an oscillatory chemical reaction, the Belousov-Zhabotinsky (BZ) reaction, occurring entirely within the gel matrix.

Yoshida and co-workers used self-oscillating gels to induce time-varying bending of beams~\cite{maeda:2007a}, transport, and peristaltic pumping~\cite{Maeda2008}.
Yashin, Balazs, and co-workers modeled self-oscillating gels with a multi-physics simulation of three-dimensional (3D) poroelastic systems coupled to the Oregonator model of BZ reaction ~\cite{dayal:2014a,yashin:2006a,chen:2011a}.
These previous studies of self-oscillating gels focused on essentially one-dimensional (1D) structures that are incapable of producing a wide range of shape changes. 

Here we present the results our of experimental and theoretical investigation of two-dimensional (2D) sheets made of self-oscillating gel.
We characterized the range of time-evolving shapes in this system, and find that 2D structures display a richer set of configurations and a wider range of BZ patterns than 1D structures.
We show that light can be used to establish or alter the pattern of the BZ reaction, and so allows the deformations to be controlled.
We quantitatively measure the evolving three-dimensional (3D) configuration of a gel and compare it to a reduced theoretical model in which the BZ phase, a single scalar field, provides the reference state of an elastic problem. The latter is expressed and solved within the formalism of incompatible elastic sheets.
This model successfully reproduces our experimental results. 

\section{Experiments} 

\begin{figure}
    \centering
    \includegraphics[width=\columnwidth]{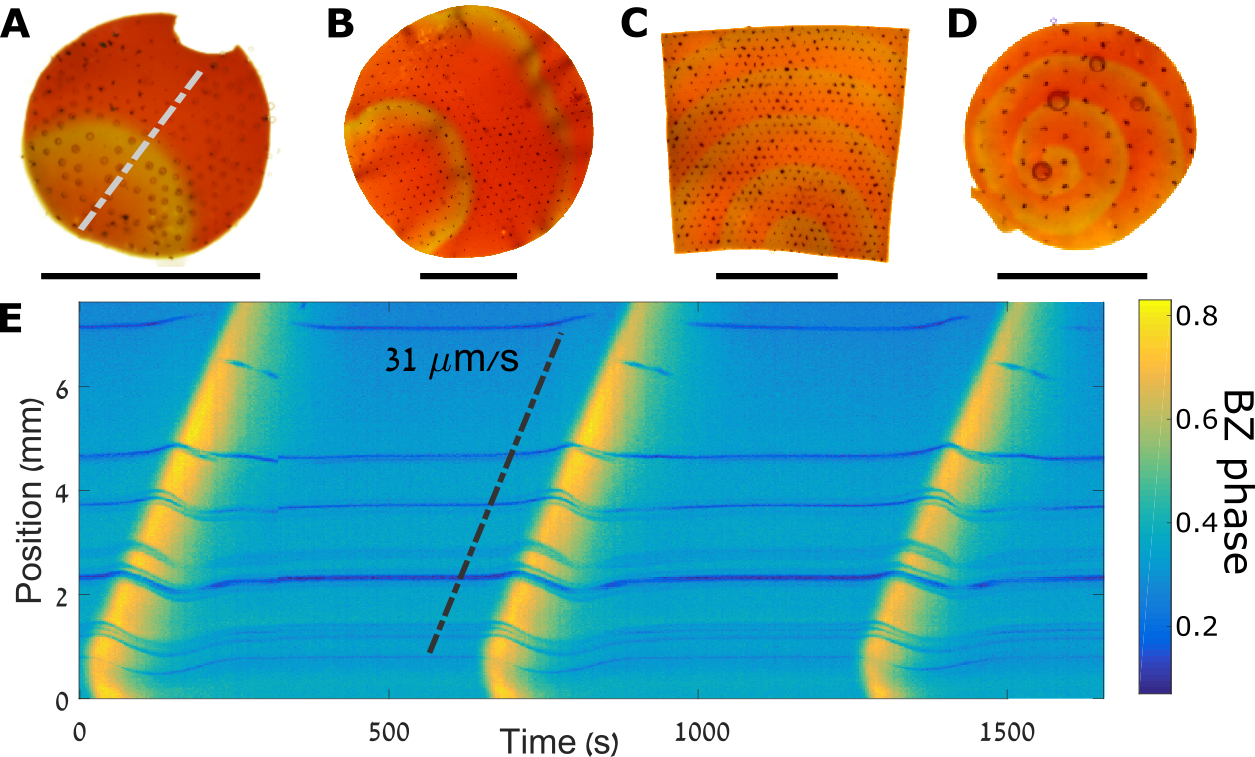}
    \caption{
        (A-D) Examples of BZ waves in a self-oscillating gel sheet for different domain shapes and chemical concentrations demonstrating the two generic patterns of the BZ reaction: target pattern (A-C) and spirals (D).
        Scale-bars correspond to 1\,cm.
        (E) Spatio-temporal map of BZ-phase along the dashed line in (A) during the passage of three BZ wavefronts.}
    \label{fig:BZexamples}
\end{figure}

\subsection{Gel preparation}
We made disc-shaped self-oscillating gels with a diameters ranging 0.5 - 5\,cm and a thickness of 500\,$\mu$m following the protocol of \citet{Maeda2008}.
We dissolved 780\,mg N-isopropylacrylamide, 81\,mg (40\,mg in the  light-control experiment) of ruthenium(II)tris-(2,2'-bipyridine) (Ru(bpy)$_3^{2+}$) (SYNTHON Chemicals GmbH, Germany) and 14\,mg of N,N'-methylenebis(acrylamide) (Sigma-Aldrich, Israel) in 2.5\,ml of methanol and 0.5\,ml of dimethyl sulfoxide.
A second solution with 27.5\,mg of 2-acrylamido-2-methylpropane sulfonic acid dissolved in 2\,ml of water was added to the mixture and purged in nitrogen for 10 minutes. 1\,ml of a 0.2\,M solution of 2,2'-azobis(2-methylpropionitrile) in toluene was added and mixed gently, and then allowed to rest until the toluene separated and rose to surface from where it was suctioned off.
The final solution was injected between two glass plates separated by a silicon-rubber gasket (Smooth-Sil 940) of the desired thickness.
The glass plates were clamped together with a purpose-built vise and placed in an oven at 60\degC\ for 20 hours.
The polymerized gels were washed in ethanol for several days. Thereafter, the gel was placed in a series of ethanol-water baths of increasing water concentration (25\%, 50\%, 75\%, 100\%) for a full day each. 

The washed gels were cut to the desired shapes with a laser-cutter and imprinted with a fine triangular grid on one side by melting spots of a thin black plastic foil onto their surface with the laser.
This grid defines a set of Lagrangian markers used in our geometrical analysis.

\subsection{Measurements}\label{sec:measurements}
\begin{figure}
    \centering
    \includegraphics[width=\columnwidth]{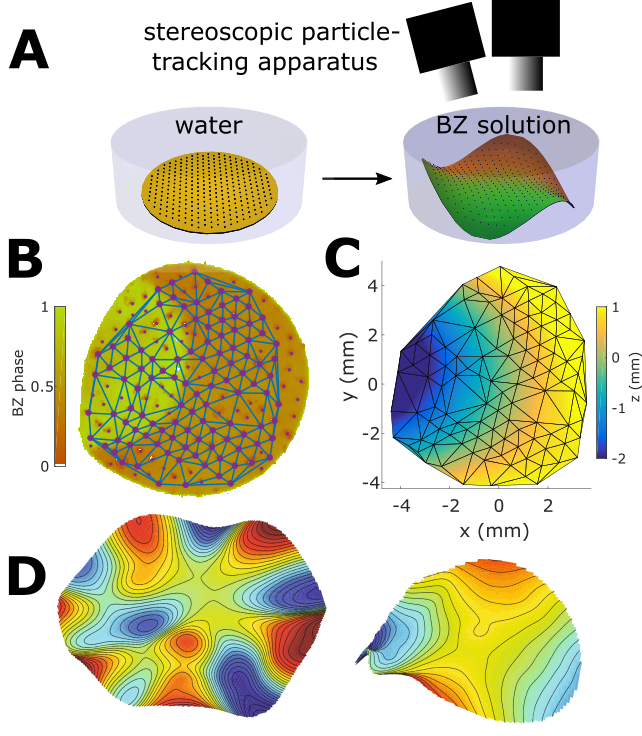}
    \caption{
        Experimental Schematic.
        (A) An initial flat self-oscillating gel was placed in a solution of the BZ reactants.
        After an induction period, green-colored oxidation traveling waves appeared in the gel that induce spatially inhomogeneous swelling and, as a result, global shape changes.
        A two-camera system was used to image grid on the surface the gel and the images were used to reconstruct the gel's three-dimensional shape and BZ-phase field.
         (B) Markers on the gels (purple dots) tracked stereoscopically with two cameras over time.
         Those markers that were successfully identified by the algorithm throughout the entire measurement (larger purple dots) were used to build a Lagrangian triangulation (blue lines) in three-dimensions.
         The BZ-phase in Lagrangian coordinates was measured at the center of each triangle.
         (C) Reconstruction of a gel's three-dimensional shape.
         The gel was adhered on its left corner to the bottom of the cell ($z=-2$\,mm) but was otherwise free.
         (D) Example of discs with different diameters 3\,cm (left) and 1\,cm (right) behaving differently for identical experimental conditions.
    }
    \label{fig:ExperimentalApparatus}
\end{figure}

Gels were placed in a temperature-controlled cell filled with 10\,ml of an aqueous solution of  0.88\,M nitric acid (HNO$_3$),  0.084\,M sodium bromate (NaBrO$_3$), and 0.062\,M malonic acid (CH$_2$(COOH)$_2$).
The solution was freshly prepared before each experiment and cooled to 20$^o$\,C before introducing the gel.
The solution contained all the BZ reactants except the catalyst Ruthenium, and so the reaction occurred solely within the gel's matrix.
After an induction period, typically  about 10 minutes long, waves of the redox state appear in the gel visible as color changes as shown in Fig.~\ref{fig:BZexamples}.
These manifest in two generic forms, shown in Fig.~\ref{fig:BZexamples}(a-d): target or spiral patterns. The wave-fronts propagate with a wave speed of approximately 30\,$\mu$m/s as shown in Fig.~\ref{fig:BZexamples}(E).
 
The wave-pattern emerges spontaneously and so its particular form is highly variable.
Particular instances may exhibit multiple centers for target or spiral waves leading to highly complicated patterns.
In the case of target patterns, the pattern self-organizes into a single center located at the edge of the gel.
Simultaneous with the emergence of the chemical waves, the gel exhibits a flapping-like motion synchronized to the reaction (see Movie S1). 

Two cameras were placed 30\,cm above the gel, one vertically above the gel in order to minimize refraction and the other tilted at 14.5$^o$ as shown in Fig.~\ref{fig:ExperimentalApparatus}(A).
The gel was illuminated from below and was photographed with both cameras simultaneously every 5 seconds.
From the two images and a triangular grid of dots printed on the top surface of the gel, we calculated the displacement field using a geometrical stereoscopy algorithm corrected for the refractive index of the solution.
Figure~\ref{fig:ExperimentalApparatus}(C) shows the shape of a gel reconstructed from such data. 
 
These reconstructions illustrate the importance of length scales.
For the same experimental conditions, two discs made from same gel but with different diameters may exhibit different displacements fields.
Figure~\ref{fig:ExperimentalApparatus}(D) shows the resulting deformation for the same target wave pattern on two different  sheets with 3 cm (left) and 1 cm (right) diameters.
The right sheet exhibits a system-spanning saddle-like shape, whereas the left example develops localized wrinkles with a well-defined length-scale. 
  
The grid points are fixed to material points, and so variations in the distance between grid points encode the local swelling/shrinking of the gel.
Given that the ratio of the gel's diameter and thickness is about 20 and so the F\"{o}ppl–von K\'{a}rm\'{a}n number is around 400, we expect the actual area is a good approximation to the reference area.
Below we show that the error is less than $\pm 4$\%.
We define the areal-growth factor $\Omega(\mathbf{x},t)$ as the ratio of the measured area to the area in the reduced state.
This typically peaks at 1.5, indicating a maximum areal swelling of 50\%.

\begin{figure}
    \centering
    \includegraphics[width=\columnwidth]{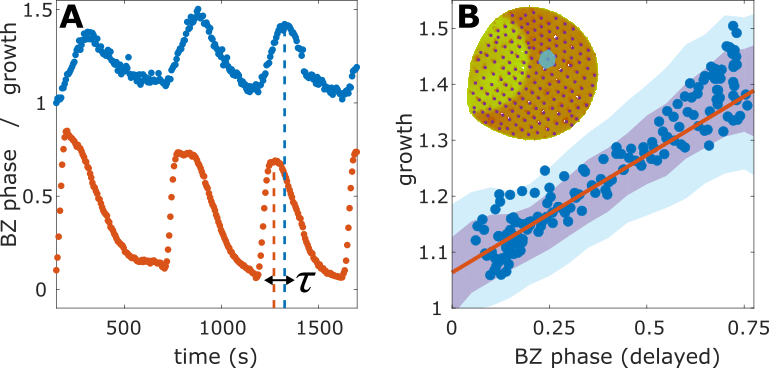}
    \caption{
        Local chemo-mechanical activity.
        (A) Average BZ phase (red) and area growth (blue) measured from the single patch indicated in (B).
        Both signals oscillate periodically, with a period of 450 seconds, and in synchrony with a time delay $\tau \approx 55$\,s.
        The magnitude of the areal swelling is roughly 40\%.
        (B) Growth as a function of the BZ phase delayed by 55\,s (blue dots).
        These data collapse to a single curve indicating that the delay is uniform in time.
        The data from all patches are represented by the colored envelopes; 75\% and 50\% of all data fall within the blue and the purple envelopes, respectively.
        The red line is linear fit to the data from all patches.
        This fit is used below to model the dependence of the growth on the BZ phase.
        Note that the fit to all data agrees well with the data from a single patch.
        Inset: the patch (teal) on the triangular grid from which the local data in (A) and (B) were collected.
    }
    \label{fig:ChemoMechActivity}
\end{figure}

Simultaneously, we measured the local phase of the BZ reaction $\phi(\mathbf{x},t)$ from the intensity of the green channel in the color images.
A comparison with the areal-growth factor in Fig.~\ref{fig:ChemoMechActivity} shows that the  signals oscillate in synchrony with a phase lag.
A cross-correlation of these signals reveals that the change in area lags behind the phase by $\tau \approx 55$\,s, a value that varies with the thickness of the gel.
Such a delay is consistent with a gradual relaxation of the actual geometry to the reference geometry due to the time scale for fluid flow through the gel’s pores.
The $\Omega(t)$ vs $\phi(t-\tau)$ in Fig.~\ref{fig:ChemoMechActivity}(B) shows that the swelling is linearly proportional to the ``delayed" phase:
\begin{equation}
  \Omega(t)= \alpha \phi(t-\tau) + \beta 
  \label{eq:constitutive}
\end{equation}
where $\alpha = 0.42  $ and $\beta = 1.06$.

Our data show that BZ waves generate local variations of the equilibrium volume of the gel.
Unlike in small gel flakes that oscillate homogeneously~\cite{Yoshida2000}, the phase of the reaction in our samples varies spatially and temporally.
Despite the uniformity of the reaction across the sheet thickness,  lateral gradients in swelling produced by the chemical waves induce local changes of the sheet’s curvature that appears on the global scale as periodic three-dimensional flapping synchronized to the reaction (Fig.~\ref{fig:BZexamples}(A), Fig.~\ref{fig:ExperimentalApparatus}(B) and Movie S1). 

\subsection{Light control}
The ruthenium catalysed BZ reaction is susceptible to light~\cite{demas:1973a,kadar:1997a}.
We found that light effect was limited for the solution described in Sec.~\ref{sec:measurements} but significant for a slightly different solution: 0.5\,M nitric acid, 0.15\,M sodium bromate, 0.01\,M sodium bromide (NaBr), and 0.05\,M malonic acid.
This solution was premixed and left to rest for 10 hours.
A gel with 50\% of the ruthenium concentration was used to reduce the required light intensity.
When the gel was immersed into the solution a spontaneous pattern emerged around the 6 hour mark after which the pattern could be changed with light. 

\begin{figure}
    \centering
    \includegraphics[width=\columnwidth]{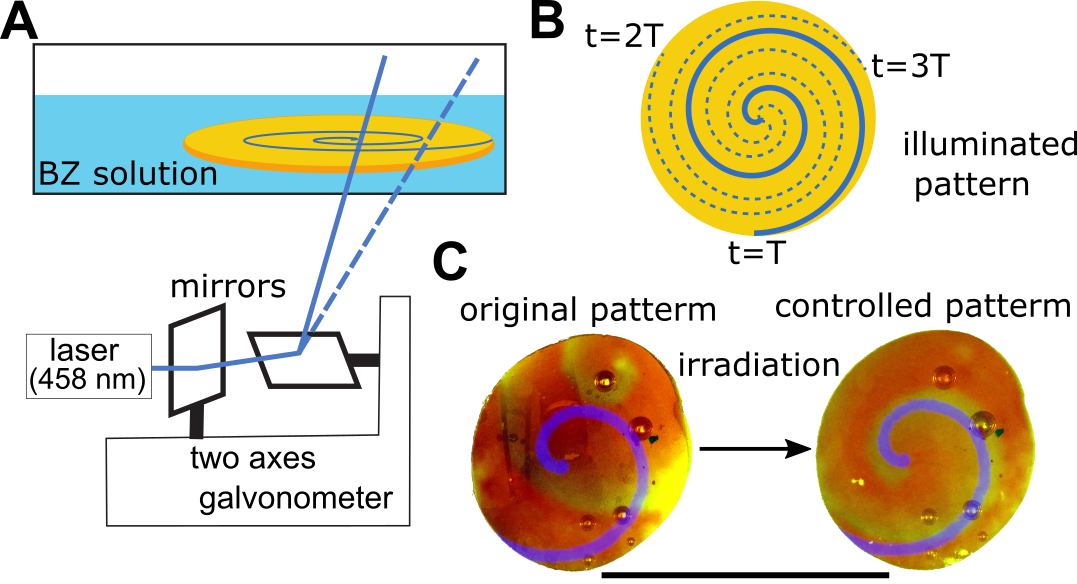}
    \caption{
        Pattern control apparatus.
        (A) A computer-controlled double-axis galvanometer apparatus is used to direct the illumination pattern of a blue laser beam.
        (B) A spiral waves was initiated by tracing out a spiral pattern with a pitch equal to that of an undriven spiral wave.  After completing each trace, the pattern was rotated to duplicate the rotational period of an undriven spiral wave.
        Each spiral trace  took 3 seconds to complete;  aproximately 100 traces were needed to initate the wagve.
        (C) The BZ pattern can be controlled (\ie suppressing the spontaneous pattern and imposing a localized spiral) by irradiating the gel with blue light.
        The blue spiral over each image is the illuminated spiral.
        Scale-bar corresponds to 1\,cm.
    }
    \label{fig:LightApparatus}
\end{figure}

The illumination was provided by a laser (wavelength 458\,nm, beam diameter 1.0\,mm, power 19.8\,mW) steered with a double-axis galvanometer apparatus as shown in Fig.~\ref{fig:LightApparatus}(A).
An Archimedean spiral with a pitch equal to that of a naturally occurring spiral wave, was repeatedly traced out on the gel for 300 seconds.
Each spiral trace took approximately 3 seconds to complete and the initial angle of the spirals was increased for each new trace to mimic the rotational period of a spiral wave (230 seconds).
An example of the results are shown in Fig.~\ref{fig:LightApparatus} and Movie S2.
Thus, the initial random pattern selected when the gel is inserted in to the solution can be replaced with any desired set of spiral wave domains.

\section{Analysis}

We applied the theory of incompatible elasticity to our system.
In this framework, the elasticity of thin sheets with non-uniform swelling is formulated in the language of differential geometry.
The resulting theories of non-Euclidean plates (NEPs) and shells~\cite{KondoBook,efrati:2009a} solves for the equilibrium configuration of the sheet in terms of the \emph{actual} metric tensor ($\tna$) and curvature tensor ($\tnb$) of the sheets midplane by extremizing the elastic energy functional
\begin{equation}
    E[\tna,\tnb]=\int d\bar{s} \left[h(\tna-\abar)^2+h^3(\tnb-\bbar)^2\right]
    \label{Eq:EnergyFunctional}
\end{equation}
given the sheet thickness $h$ and the \emph{reference} metric ($\abar$) and curvature ($\bbar$) tensors dictated by the swelling field. The reference metric field encodes the local equilibrium distances between points in the midplane. Deviations of the actual metric from the reference one (\ie  $\tna\neq\abar$) leads to stretching energy, which is the first term in Eq.~\ref{Eq:EnergyFunctional}. The reference curvature encodes gradients of equilibrium distances perpendicularly to the midplane. Deviations of the actual curvature from the reference one (\ie $\tnb\neq\bbar$) leads to bending energy, which is the second term in Eq.~\ref{Eq:EnergyFunctional}.
The resulting shape $\mathbf{r}(\mathbf{x})$ can be  reconstructed by integrating the shape operator $\mathcal{S}^i_k\equiv \tna^{ij} \tnb_{jk}$.
  
Clearly, Eq.~\ref{Eq:EnergyFunctional} is at a minimum when both $\abar=\tna$ and $\bbar=\tnb$.
However, it is often impossible to construct such a shape because $\abar$ and $\bbar$ are independent fields, dictated by the swelling profile. In contrast, $\tnb$ and $\tna$ are derived from a configuration, thus are related by Gauss's \emph{Theorema Egregium}. If the swelling profile is such that $\abar$ and $\bbar$ do not satisfy Gauss's \emph{Theorema Egregium}, the sheet is incompatible and it cannot have a stress-free configuration.
In such cases, the actual geometry emerges from a competition between the bending term, proportional to $h$, and stretching term, proportional to $h^3$.

Solving Eq.~\ref{Eq:EnergyFunctional} is generally a difficult problem, but for thin sheets it is  useful to first perform a purely geometrical analysis ($h \rightarrow 0$ limit) followed by - if necessary - a correction via a mechanical analysis.
In the limit $h\rightarrow 0$ the sheet is bendable but unstretchable, and therefore  $\abar \approx \tna$ and the \emph{actual} Gaussian curvature is completely determined by the \emph{reference} metric.
The patchwork of positive and negative curvature domains is often a good rough guide to the shape of the sheet, but minimizing the total mean curvature while maintaining $\tna = \abar$ is necessary to uniquely specify the shape. 

As shown in Appendix~\ref{appdx:isotropy} the deformation field for a self-oscillating gel is locally isotropic.
Furthermore, since our gels are very thin, the BZ-phase is uniform across the thin dimension.
Therefore, the reference curvature vanishes ($\bbar=0$) and the reference metric is always conformal to a flat metric:
\begin{equation}
        \abar(\mathbf{x},t)=\Omega(\mathbf{r},t)
    \begin{pmatrix}
        1&0\\
        0&1
    \end{pmatrix}=
    \exp \left(\sigma(\mathbf{x},t)\right)
    \begin{pmatrix}
        1&0\\
        0&1
    \end{pmatrix}
\end{equation}
where $\sigma$ is the logarithm of the areal swelling factor $\Omega$. For this metric, the reference Gaussian curvature reduces to (see~\cite{oneillbook} page 297):
\begin{equation}
    \Kbar(\mathbf{x},t)=-\frac12\Delta_{LB}\sigma(\mathbf{x},t)
    \label{eq:ref_curv}
\end{equation}
where $\Delta_{LB}\equiv e^{-\sigma} \nabla^2 $ is the Laplace-Beltrami operator (the curved-space Laplacian) for the reference metric (see Appendix~\ref{appdx:derivatives}).
We compute $ \Kbar(\mathbf{x},t) $ from the BZ field using Eqs.~\ref{eq:ref_curv} and~\ref{eq:constitutive}. 

\begin{figure}
    \centering
    \includegraphics[width=\columnwidth]{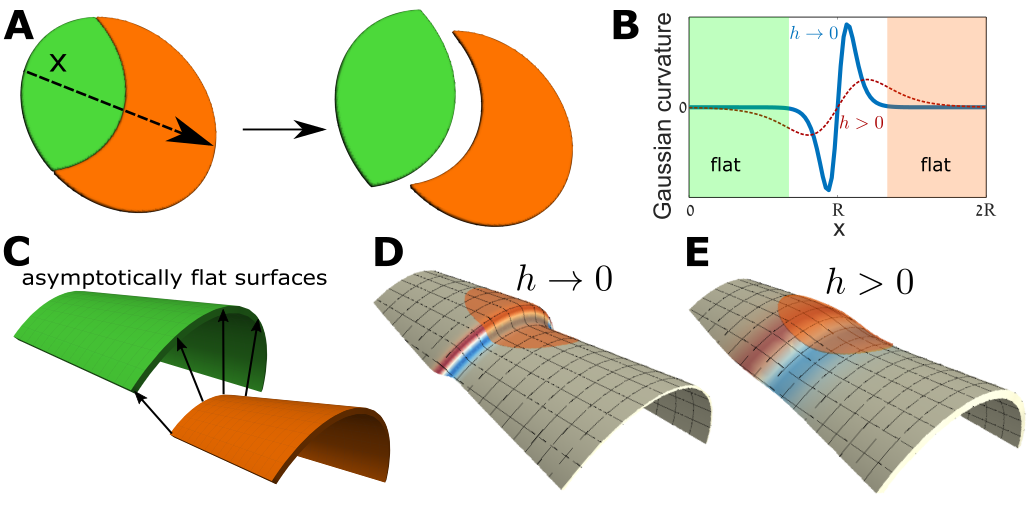}
    \caption{
        Approximate reconstruction of the surface from the Gaussian Curvature.
        (A) Illustration of the intrinsic geometry a gel exhibiting a BZ pattern similar to Fig.~\ref{fig:ChemoMechActivity}.
        The gel is comprised of two incompatible flat segment that are glued together at the circular front.
        Such intrinsic geometry clearly prohibits a flat unstretched configuration.
        (B) The Gaussian curvature along the marked line vanish everywhere except for the large oscillation at the front.
        (C) An ansatz of a surface of revolution together with this form of Gaussian curvature lead to a shape made of two conical surfaces glued together by a smooth region.
        (D) An illustration of such embedding of the reference metric.
        Since the front width is very small, the amplitude of the Gaussian curvature near the front is very large resulting in a localized bending energy.
        The orange circles corresponds to the size of the experimental gel.
        (E) Finite thickness effects are expected to spread the Gaussian curvature over a wider area reducing the bending energy.
        Such a correction to the thin limit maintains the intrinsic geometry far from the front while reducing the bending energy at the front.
    }
    \label{fig:bottle}
\end{figure}

The resulting field provides useful, albeit incomplete, information about the 3D shape of the sheet.
For example, in the case of a single propagating BZ front, such as the one shown in Fig.~\ref{fig:ExperimentalApparatus}B, the reference Gaussian curvature field is zero on either side of the front and oscillates from zero to positive to negative and back to zero across the front.
Qualitatively, this corresponds to a surface of revolution consisting of two conical segments smoothly connected as shown in Fig.~\ref{fig:bottle}.

\begin{figure}
    \centering
    \includegraphics[width=\columnwidth]{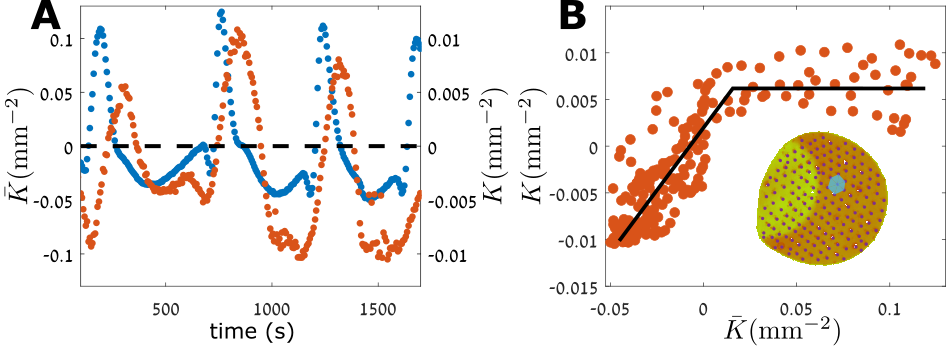}
    \caption{
        Analysis of Gaussian curvature.
        (A) Time-dependence of the reference (blue) and actual (red) Gaussian curvatures in a small patch.
        The peaks in $\Kbar$ are due to sharp changes in the reference metric.
        Note that the oscillations in $K$ are weaker than in $\Kbar$, and that $K$ lags behind $\Kbar$ because of the lag in the growth (see Fig.~\ref{fig:ChemoMechActivity}).
        (B) Actual versus reference Gaussian curvature. The actual curvature is plotted with the delay $\tau$.
        Inset: the teal patch indicating the location from which the data in (A) and (B) were collected (the same as in Fig.~\ref{fig:ChemoMechActivity}).
    }
    \label{fig:gaussian_curvature}
\end{figure}

The actual Gaussian curvature $K(\mathbf{x},t)$ was extracted from the three-dimensional surface measurements.
The actual and reference curvatures for a single BZ pulse are plotted in Fig.~\ref{fig:gaussian_curvature}.
The asymmetry between positive and negative values of both the reference and actual curvatures arise from the asymmetry of a BZ wave (see \eg Fig.~\ref{fig:ChemoMechActivity}A) which is rapidly changing at the front and slowly varying in the tail.
Figure~\ref{fig:gaussian_curvature} shows that the actual and reference Gaussian curvature exhibit similar temporal pattern, but differ in magnitude.
For small $|\Kbar|$, $K$ and $\Kbar$ are linearly proportional (\ie  $K= \frac1{10} \Kbar$); for $\Kbar>0.01$\,mm$^{-2}$, $K$ is constant independent of $\Kbar$ (\ie $K \approx 0.007$ \,mm$^{-2}$).
  
The difference in magnitude between $K$ and $\Kbar$ together with the cutoff in $K$ reveal that the system is not in the asymptotic limit $h\rightarrow  0$  where $K=\Kbar$ everywhere.
On the other hand, the qualitative similarity of the temporal data show that the system is still sufficiently thin that the geometrical analysis provides the correct qualitative picture. 

The competition between stretching energy and bending energy for $h\neq 0$ softens the reference Gaussian curvature so that the actual Gaussian curvature is smeared out as seen in Fig.~\ref{fig:bottle} \& \ref{fig:gaussian_curvature}.
The introduction of a new length-scale (i.e. the width of the transition in $\abar$) allows for intermediate regimes that lie between the thin limit (stretching dominated) and the thick limit (bending dominated)\cite{kim:2012b,moshe:2013a}.
In this regime, $K$ deviates from $\Kbar$ in the rapidly changing region; its amplitude is lower, but the decay length-scale is longer. 
Inserting values from our experiments into this theory yields $\Kbar/K \approx 10$ (see Fig. 3B in~\cite{moshe:2013a}), consistent with the observation in Fig.~\ref{fig:gaussian_curvature}(B). 

Our data indicate that oscillations in the BZ field control the evolution of the three-dimensional configuration, but not in a purely-geometrical manner.
Thus, the system needs to be treated as a non-Euclidean elasticity problem by finding the equilibrium 3D configuration of a plate for a given reference metric~\cite{efrati:2009a}.
Such problems have been extensively studied and it is known that the 3D configuration is set by a competition between stretching and bending.
As in the case of constrained flat sheets~\cite{Davidovitch2011}, for the same reference metric there are bifurcations in the equilibrium shape controlled by dimensionless groupings, such as the ratio of thickness to lateral extent~\cite{efrati:2009b,armon:2011a}.
We observe such a bifurcation in our system (\eg Fig.~\ref{fig:ExperimentalApparatus}(D)).
The deformation caused by a circular BZ front can produce a system spanning saddle-like shape on a small thick disc (Fig.~\ref{fig:ExperimentalApparatus}D right, Movie S1) or a wrinkled shape on a large thin disc (Fig.~\ref{fig:ExperimentalApparatus}D left, Movie S4). From here  onward we concentrate on the saddle-like regime.

\section{Numerical analysis}
Since an analytical solution to the full elastic problem is unavailable, we instead used numerical solutions to study the deviations from the $h\rightarrow 0$ limit.
Using a non-Euclidean finite-element code, we computed the shape of a disk with a radius of 5\,mm and uniform thicknesses 0.4, 0.5, 0.6, or 0.8\,mm subject to a non-uniform growth of peak amplitude $\Omega_o$ along a circular front of width $\delta$ centered at various distance $d$ from the edge.
Since the typical widths were of the same order-of-magnitude as the thickness, we used a fine grid of over 4000 elements to precisely capture the elastic response.
The specific functional form of the front is inconsequential and for simplicity was chosen to be $\Omega = 1+ \frac12 \Omega_o\left[1+\tanh\frac{d-\|\mathbf{r}-\mathbf{r_o}\|}{\delta}\right]$ with $\delta =0.25$\,mm  and $\Omega_o=0.5$ chosen to match experimental values.
By varying the distance of the front from its origin, we simulated the propagation of the BZ front: for each value of $d$ we computed the equilibrium configuration and this sequence of configurations was taken as an approximation of the time-dependent geometry. 

\begin{figure}
     \centering
     \includegraphics[width=\columnwidth]{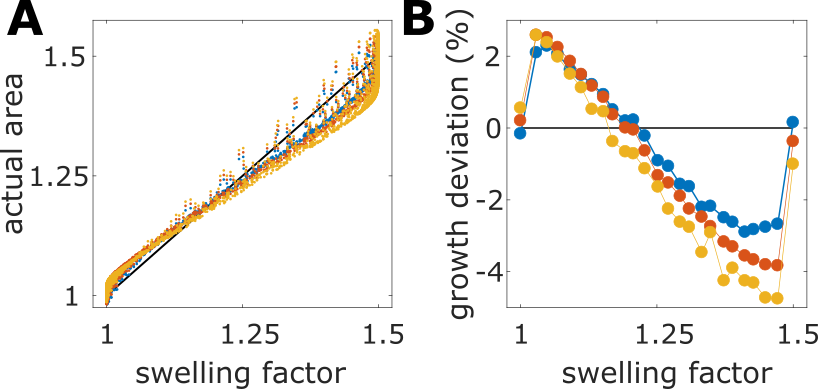}
     \caption{
         Deviations of the actual area from the swelling field from numerical simulations.
         (A) The actual versus reference area for three different thicknesses (0.4, 0.6 and 0.8 mm in blue, red and yellow, respectively).
         (B) The relative error of the actual area from the reference area (normalized by the swelling factor).
         As expected from elasticity, the error increases with  thickness, is negative in the swollen regions and positive in the shrunk regions, and less than 4\% for the relevant thicknesses.
         }
     \label{fig:Deviations}
 \end{figure}

We first estimated the error introduced by using the actual area instead of the reference area in the relationship between the BZ phase and the growth.
Experimentally we can only measure actual properties.
Hence, in the calibration of the relationship between swelling and the BZ phase, we compared the phase to the actual area under the assumption that the latter is a good approximation of reference area.
While this assumption is exact in the $h\rightarrow0$ limit where $\tna=\abar$, our analysis indicates the presence of finite-thickness effects (see Fig.~\ref{fig:gaussian_curvature}).
The numerical computations show, as expected from the {high} F\"{o}ppl–von K\'{a}rm\'{a}n number, that the difference between the actual and the reference area is less than 4\% and largely localized to regions where the metric is changing rapidly (see Fig.~\ref{fig:Deviations}).
This justifies our interpretation of Fig.~\ref{fig:ChemoMechActivity} as representing the connection between the BZ phase and $\Omega$.  

\begin{figure}
     \centering
     \includegraphics[width=\columnwidth]{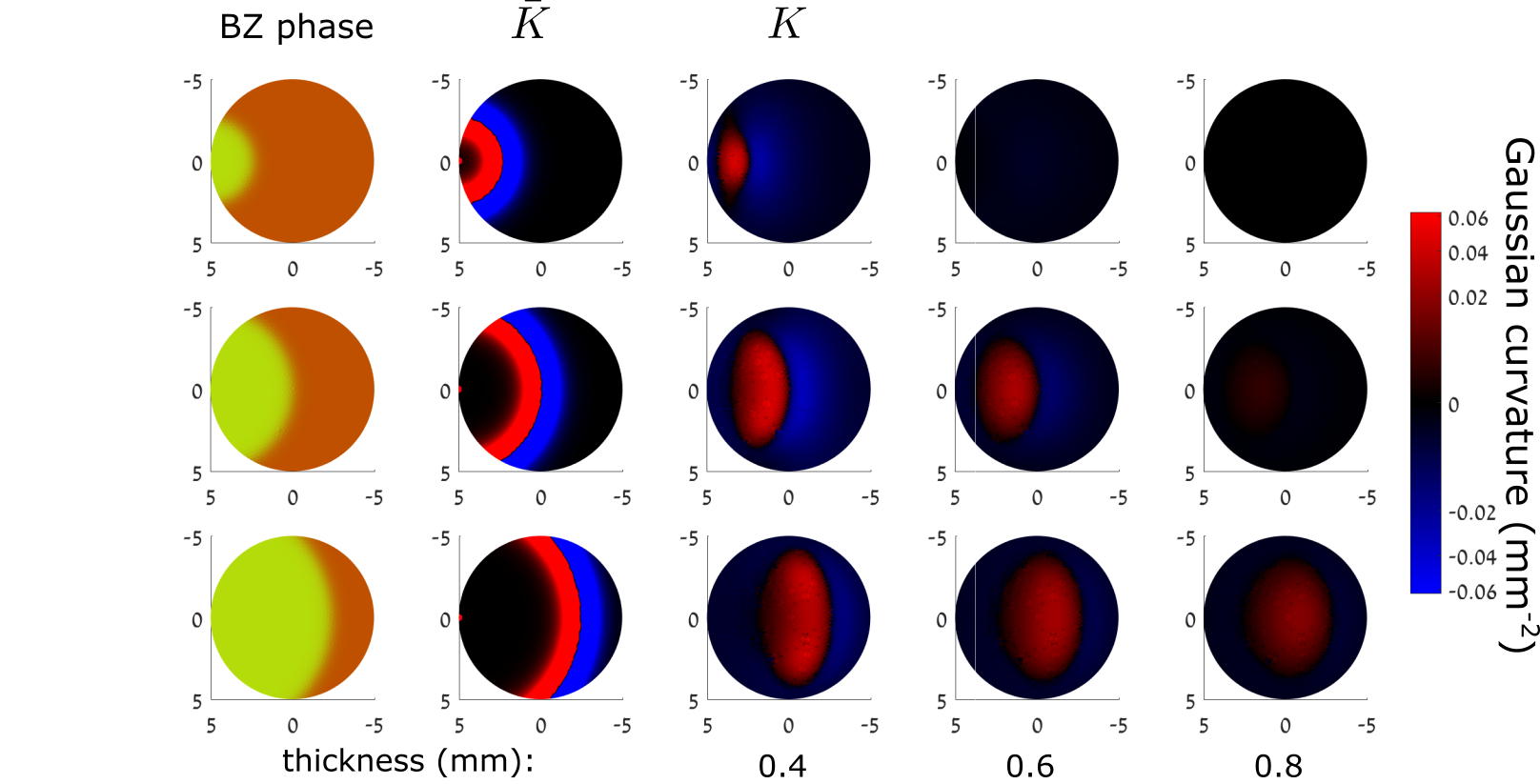}
     \caption{
         Numerical results for finite-thickness effects.
         Column 1: BZ phase input to simulations (top view) where each row corresponds to a later time.
         Column 2: reference Gaussian curvature computed from the BZ field using Eq.~\ref{eq:ref_curv}.
         Column 3 - 5: actual Gaussian curvature for thicknesses of 0.4, 0.6, 0.8 mm  from left to right (values are indicated below the panels).
         The changes in the actual Gaussian curvature as the front propagates determines the evolution of the three-dimensional shape.
         Note: the color is scaled nonlinearly in order to present the reference and actual values on the same scale.
     }
     \label{fig:FiniteThicknessEffect}
 \end{figure}

While the differences between the reference and actual area are small, the same is not true of the Gaussian curvatures.
In our experiments we measure a difference of up to an order of magnitude where the metric varies rapidly.
Figure~\ref{fig:FiniteThicknessEffect} shows numerical simulations for various thicknesses.
The comparison of the reference and the actual Gaussian curvature for different thicknesses shows that for thicker sheets the peak magnitude of the actual Gaussian curvature decreases and the width transition area increases due to the increased cost of bending a thicker sheet.
Thus, even though the values of the metric tensors are similar, the broader width of the transition area leads to a large change in the actual curvature because the curvature involves a second derivative. 

\begin{figure}
    \centering
    \includegraphics[width=\columnwidth]{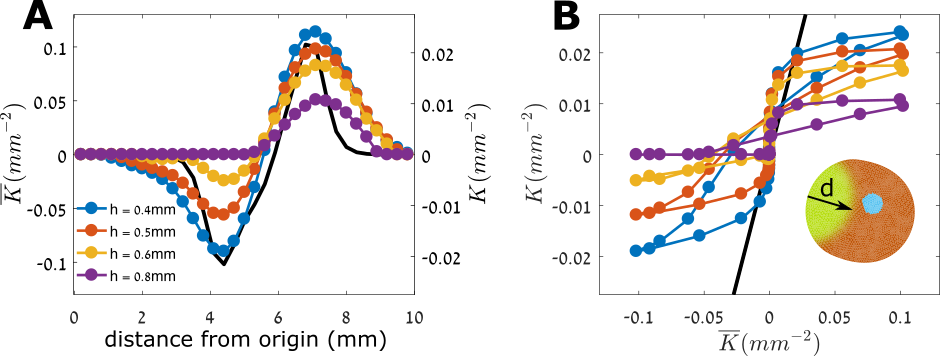}
    \caption{
        (A) Reference curvature versus the distance from the origin of the propagating front from measurements (black) and for different sheet thicknesses from numerical simulations (colored).
        (B) Actual versus reference Gaussian curvature from simulations for finite thickness sheets.
    }
        \label{fig:gaussian_numerical}
\end{figure}

\begin{figure}
     \centering
     \includegraphics[width=\columnwidth]{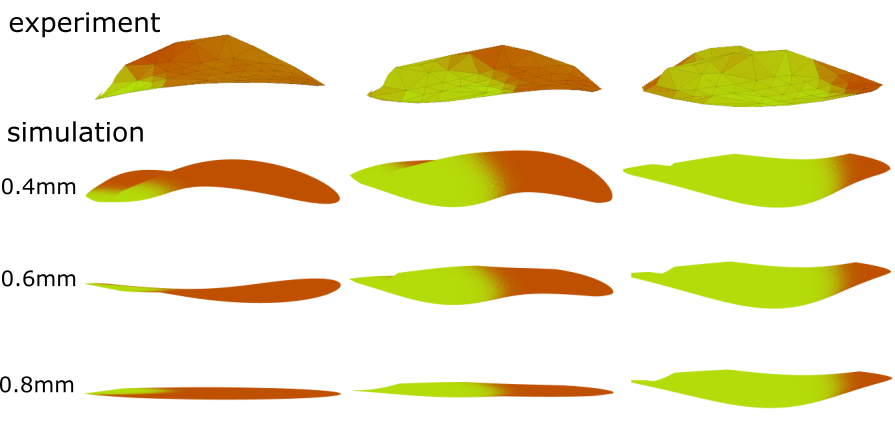}
     \caption{
         Experimental and numerical shape-evolution of a BZ gel disk.
         Top row: side view of shapes measured during one period.
         Rows 2 - 4: side views of shapes from finite-element simulations for different thicknesses of 0.4, 0.6, 0.8\,mm (top to bottom).
         The spatial variation of the BZ phase was selected to match the characteristics of the fronts observed in experiments (amplitude 50\%, width 250\,$\mu$m and diameter 10\,mm).  
     }
     \label{fig:ShapeEvol}
 \end{figure}

Next we compared the measured curvature with the computed curvature for various thicknesses.
As shown in Fig.~\ref{fig:gaussian_numerical}(A), the profiles are qualitatively similar, including the changes in the curvature sign across the shape’s evolution, but that the profile for a thickness of 0.4\,mm most closely matches quantiatively the experimental data.
Furthermore,  Fig.~\ref{fig:gaussian_numerical}(B) shows that actual curvature displays a similar cutoff for a thicknesses of 0.4\,mm. 
Finally, we computed the shape for various thicknesses.
These data are compared with experiments in Fig.~\ref{fig:ShapeEvol} and again, the quantitative agreement is best for  a thickness of 0.4\,mm.

\section{Discussion}  
Our experiments show that the reference metric is set by the phase of the BZ reaction and that the reference curvature is zero.
Thus, the entire effect of the chemical field can be encoded in the reference metric, $\abar$, or equivalently in the reference Gaussian curvature field.

This work provides the first realization of a synthetic autonomous multi-axial deformable sheet.
The sheet “metabolizes” chemical energy to produce mechanical energy, \ie periodic three-dimensional shape changes.
The underlying physics is well described by the theory of incompatible elastic sheets, where the entire effect of the chemical field can be encoded in the reference metric, $\abar$.
The theory successfully describes the shape selection of the sheet, which is governed by a competition between stretching and bending.

Other processes, such as hydrodynamics and feedback, may affect the evolving shape, mainly by selecting different embedding of the reference geometry.
Hydrodynamic effects, neglected here, are significant when the sheets are large and thin and when the evolution is fast.
Additional dimensionless parameters are needed to include this effect and may potentially lead to a richer configuration space.  We also neglected possible feedback of the deformation on the propagation of BZ wave~\cite{Miller2018} that may also be significant in thinner sheets.

The concepts presented in this work - the autonomous conversion of planar deformations into curvature fields, the cutoffs in time and curvature - are general.
They apply to many different systems, such as plates with non-flat background (non-oscillating) geometry~\cite{klein:2007a,kim:2012a} or the actuation of evolving spontaneous curvature due to gradients across the thickness, and are likely to apply to living cells~\cite{Nitsan2016,Park2016} or be realized by different synthetic materials such as nematic elastomers~\cite{Gelebart2017,McConney2013}.
We expect our results will stimulate new approaches to realizing autonomous soft machines. 

\appendix
\section{Isotropy of local growth field\label{appdx:isotropy}}
In order to calculate the reference Gaussian curvature, one must know the tensorial form of the reference metric and its dependence on the local growth rule.
A  local growth field generically transforms circles into ellipses and is characterized by the areal growth factor, and the eccentricity and orientation of the ellipses. The expression we used for the reference Gaussian curvature (Eq.~\ref{eq:ref_curv}) holds for isotropic deformation only.
The data presented below shows that indeed the growth field of self-oscillating gels is isotropic, \ie circles are transformed into circles, thus justifying using Eq.~\ref{eq:ref_curv}.

\begin{figure}
     \centering
     \includegraphics[width=\columnwidth]{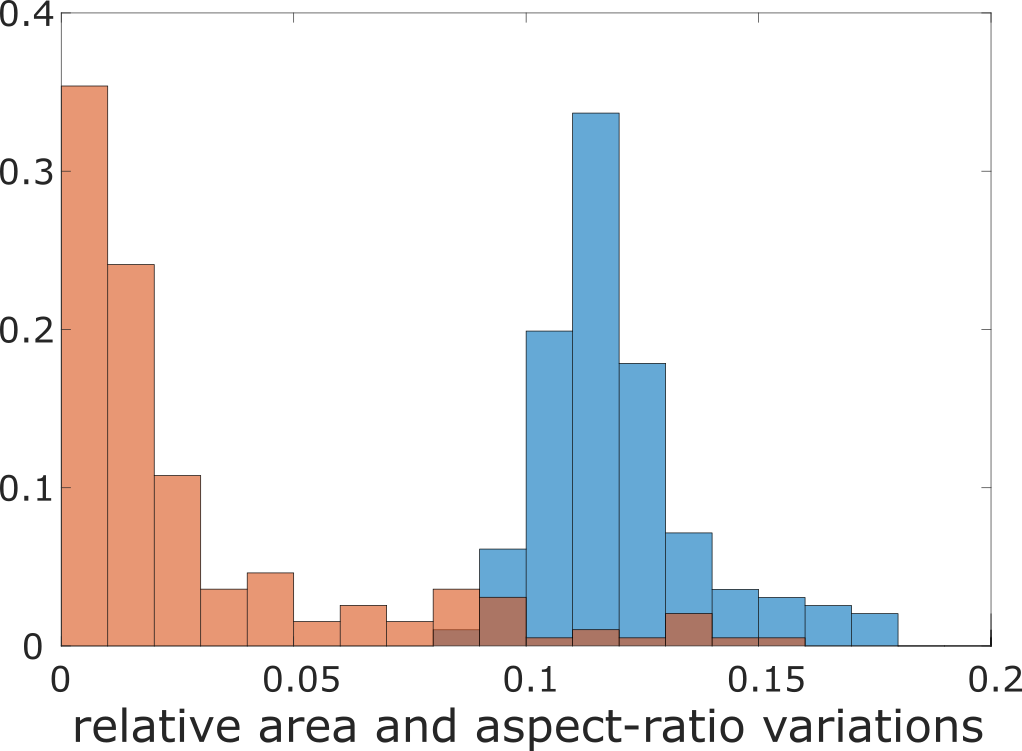}
     \caption{
         Growth Isotropy.
         A histogram of the relative fluctuations, defined as the variance normalized by the mean, in the aspect-ratio (orange) and the area (blue) of each triangle.
         The tendency for small values of the aspect-ratio fluctuations (lower than 2\%) are indicators of isotropic growth.
     }
     \label{fig:GrowthIsotropy}
 \end{figure}

We define the aspect-ratio of the triangles in the mesh ($\bigtriangleup abc$) as 
\begin{equation}
S \equiv abc/\left((a+b-c)(c+a-b)(b+c-a)\right)
\end{equation}
and its fluctuations as $\delta S\equiv\sqrt{\mbox{Var} [\Delta S]}/\langle\Delta S\rangle$ where $\Delta S$ is the change in $S$ from its initial undisturbed state.
Note that for an isotropic dilation $\Delta S = 0$.
The histogram in Fig.~\ref{fig:GrowthIsotropy} shows that $\delta S$ is strongly peaked at zero and the average is 0.03, unlike the fluctuations of the area which are centered around an average of 0.125.
The fact that the fluctuations of the aspect ratio are negligible compared to ones of the area confirms that the growth is isotropic Fig.~\ref{fig:GrowthIsotropy}.

\section{Using the reference geometry to calculate spatial derivatives\label{appdx:derivatives}}
In any formalism with both reference and actual geometries, the relevant geometry must be defined for each integration or differentiation.
In our two-dimensional non-Euclidean models, the integration in Eq.~\ref{Eq:EnergyFunctional} was defined with respect to the reference geometry since this is the geometry input to the model.
However, this choice is only significant for higher orders in the thickness and therefore irrelevant for the bending and stretching terms.
In this work, we treat the BZ dynamics as given and study the elastic response.
However, a full model for this problem that also incorporates the BZ-phase dynamics may require a carefuller definition of the spatial derivatives in the reaction-diffusion equation. 

\bibliographystyle{apsrev4-2}
\bibliography{BZgelBib4.bib}

\end{document}